\documentstyle[aas2pp4]{article}

\newcommand{\di}{$\Delta I$\ }
\newcommand{\ddi}{$\Delta I$}
\newcommand{\dpol}{$\Delta P$\ }
\newcommand{\ddpol}{$\Delta P$}

\begin{document}

\title{Blobs in Wolf-Rayet Winds: \\Random Photometric and Polarimetric Variability}
\author{Cl\'audia~V. Rodrigues\altaffilmark{1}}
\affil{Instituto Nacional de Pesquisas Espaciais/MCT\\Av. dos Astronautas,
1758\\12227-900 - S\~ao Jos\'e dos Campos - SP\\Brazil\\
e-mail: claudia@das.inpe.br}
\authoremail{claudia@das.inpe.br}
\and
\author{A.~M\'ario Magalh\~aes}
\affil{Instituto Astron\^omico e Geof\'\i sico, Univ. de S\~ao Paulo\\
Av. Miguel St\'efano, 4200\\04301-904 - S\~ao Paulo - SP\\Brazil\\
e-mail: mario@argus.iagusp.usp.br}
\authoremail{mario@argus.iagusp.usp.br}
\altaffiltext{1}{formerly at Instituto Astron\^omico e Geof\'\i sico, Univ. de S\~ao
Paulo}

\lefthead{Rodrigues \& Magalh\~aes}
\righthead{Blobs in Wolf-Rayet winds}

\begin{abstract}

Some isolated Wolf-Rayet stars present random variability in their
optical flux and polarization.  We make the assumption that such
variability is caused by the presence of regions of enhanced density,
i.e. blobs, in their envelopes. In order to find the physical characteristics
of such regions we have modeled the stellar emission using  a Monte Carlo code
to treat the radiative transfer in an inhomogeneous electron scattering
envelope.  We are able to treat multiple scattering in the regions of
enhanced density as well as in the envelope itself. The finite sizes of
the source and structures in the wind are also taken into account.
Most of the results presented here are based on a parameter study of
models with a single blob. The effects due to multiple blobs in the envelope
are considered to a more limited extent. Our simulations indicate that the
density enhancements must have a large geometric cross section in order to
produce the observed photopolarimetric variability.  The sizes must be of the
order of one stellar radius and the blobs must be located near the base of the
envelope. These sizes are the same inferred from the widths of the sub-peaks
in optical emission lines of Wolf-Rayet stars.  Other early-type stars show
random polarimetric fluctuations with characteristics similar to those
observed in Wolf-Rayet stars, which may also be interpreted in terms of a
clumpy wind.  Although the origin of such structures is still unclear, the
same mechanism may be working in different types of hot stars envelopes to
produce such inhomogeneities.

\end{abstract}


\section{Introduction}

Our current understanding of hot star envelopes pictures them as
inhomogeneous and non-steady structures (Moffat et al. 1994; Wolf,
Stahl \& Fullerton 1999; Magalh\~aes \& Rodrigues 1999). In
particular, Wolf-Rayet (WR) stars can present random variations in
broad band flux, polarization and also in spectral line profiles.
The broad band fluctuations can reach 10\% in flux (Antokhin et
al. 1995; Marchenko et al. 1998) and 0.5\% in polarization
(St.-Louis et al. 1987; Drissen et al. 1987; Robert et al. 1989;
Moffat \& Robert 1991). The observed changes in the spectral line
profiles can be divided in two types. One of them is the small
moving bumps which appear superposed on optical emission lines
(Robert 1994). Moffat \& Robert (1992) suggested that they are
related to localized enhancements of material, i.e. blobs, in WR winds.
The other type of variation seen in WR stars are discrete absorption
components (DACs; Prinja \& Smith 1992). In other hot stars, such as OB stars,
DACs are also observed and seem to be connected to the rotation
period of the star (Kaper \& Henrichs 1994). The DACs in WR stars
may then be not related to the random broad band fluctuations,
since in other hot stars these features seem to be periodic. 
Some kind of periodic line variability is seen in a few isolated
WR stars - WR~1, WR~6 and WR~134 (see Lepine, Eversberg \& Moffat 1999
and references therein). However, for the stars having polarization
measurements (WR~6 and  WR~134), the position angle of the polarization is
relatively constant, a behavior different from that seen in most of the
isolated WR stars which present randomly varying position angles. Therefore
the isolated WR stars which present some kind of periodic behavior seem to also
have envelopes with a time-independent anisotropy, in contrast to the
randomly variable stars. Less direct evidence for clumpy WR winds can be
obtained from the fits of line wings (Hamann \& Koesterke 1998; Hillier 1984).

There is evidence that the random variability in the three types of
observations (flux, polarization, and spectra) are different facets of the
same phenomenon: (1) these variability types are correlated (Robert
1994); (2) the time scales are the same, ranging from hours to days
(\cite{mr91}). Robert (1994) has also shown that the spectral variations
are caused by localized density enhancements whose frequency of
occurrence increases for smaller masses. The size distribution of these
spectral features is also addressed in Lepine \& Moffat (1996, 1999). The
velocity dispersion in these features suggests a physical dimension around one
stellar radius for such blobs (Robert 1994).

Theoretical results indicate that hot star envelopes are indeed not
expected to be homogeneous. In his review, Owocki (1994) has suggested
that early-type star winds can have two types of instabilities. There is
a small-scale instability, which is intrinsic to a radiative wind and can
be responsible for the bumps observed in the WR emission lines (see also
Gayley \& Owocki 1995). On the other hand, external processes, such as
rotation, binarity, and photospheric processes, can induce instabilities
on large scales. While this second mechanism may produce the DACs, the
former may be associated with the so-called blobs in WR winds.

The determination of the physical properties of the inhomogeneities in
WR wind constrains the theoretical models for their origin as well as
the models for the mechanism producing the wind itself. In the last
few years, some studies have tried to model the random polarimetric
variations in hot stars (Fox \& Henrichs 1994; Brown 1994; Brown et
al. 1995; Code \& Whitney 1995; Richardson, Brown \& Simmons 1996).
All these models take into account only the region of enhanced density,
consequently the envelope itself is not considered as an element to
modify the radiation coming from the star or that scattered by a blob.
Most of them are restricted to point-like regions and single scattering
treatment, the exception being the numerical Monte Carlo treatment
of Code \& Whitney (1995). Studying the random photopolarimetric
variability in WR stars, Brown (1994) has concluded that the density
enhancements may be originated at the stellar surface and not be the
result of a mass redistribution process in the wind, such as turbulence. In
a subsequent paper, Brown et al. (1995) have pointed out that the
fractional variation in flux and polarization of the continuum light
must be of a similar magnitude if they are produced by scattering of
starlight, a fact which is not observed; as mentioned early in this
paper, the fluctuations in broad band flux are much larger than the
polarization. Richardson et al. (1996) have revisited this problem considering
an envelope with many blobs and concluded that the observed ratio between the
photometric and polarimetric variations, $\Delta I$/$\Delta P$, cannot be
reproduced by any number of inhomogeneities. They have suggested that the
emission from blobs might be an important contributor to the total flux and
this would bring the models closer to the observations. In order for this be
valid, the blobs have to be very dense.

In this work, we investigate whether blobs can explain the changes in
flux and polarization observed in WR stars and what would their resultant
physical characteristics be, as the size and distance from the star. We
have included, besides scattering, extinction as a mechanism to produce
flux variations. We have used a Monte Carlo code that does not have the
limitations of single scattering or point-like sources. Unlike earlier
works, scattering in the envelope itself is also considered. This is important
since WR stars have optically thick envelopes.  In the following section, we
introduce the main features of our model. In Section 3, we describe how the
resulting broad band variations depend on the input parameters for a model
with a single blob. In this section, some results for envelopes with more than
one blob are also presented. The model results are applied to WR envelopes in
Section 4. The applicability of the model for other hot stars and a summary
are then presented.

\section{The model}

\subsection{Model assumptions}
 
We assume that the observed broad band random variability in WR
stars is caused by the presence of regions of enhanced density which we
will call blobs. Here we are not interested in explaining their
origin: our goal is just to constrain the physical characteristics that the
blobs may have in order to explain the observed quantities.  We have
solved the radiative transfer in an electron scattering envelope using
the Monte Carlo code described in \cite{rod97}. This code has already
been used to study the light curve and polarization of WR binary stars
(Rodrigues \& Magalh\~aes 1994; Rodrigues 1997). The basics of a
Monte Carlo code to simulate radiative transfer is described in \cite{cw95}.

The blobs have been assumed to be spherical and immersed in a spherical
envelope. The density law of the envelope can be chosen among many
analytical expressions: constant, inversely proportional to the
square of the distance, or following a beta-type law for the wind
velocity in radiative driven winds $[v(r) \propto \left( 1 - R_\star/r
\right)^\beta)]$, among others.  The density within the blob follows
the same law as the envelope, but it is multiplied by an arbitrary
factor which introduces a density enhancement. For instance, if this
factor is equal to one, the envelope is isotropic (no blobs exist at
all). Therefore, our model does not represent density enhancements caused
by redistribution of matter in the envelope. It is consistent with Brown's
(1994) statement that polarization is not caused by a process with such
a characteristic.  In addition, we are able to treat an arbitrary number
of blobs ($N_{bl}$). The only source of radiation is an extended, spherical central
source, i.e., the envelope and the blobs do not emit any radiation.
The use of the Monte Carlo method has allowed us to consider
optically thick envelopes characteristic of WR winds. The last point
is an important difference in comparison with previous models, which have
considered no envelope at all.

The number of photons ($N$) of each simulation was either $3 \times 10^7$ or 
$3 \times 10^8$ for optically thin and thick envelopes, respectively.  The
exiting photons were classified among 400 bins of identical solid angle. The
light emitted by the source was considered to be isotropic and not polarized.
Some of the photons can hit the source. Initially, we have considered
they were absorbed by the source. However, reemission of these photons may in
principle have an effect on the results\footnote{We wish to thank J. Bjorkman
for pointing that out and discussing it with us.}. We have then modified our
code in order to include the backwarming of the stellar photosphere. This was
done by isotropically reemitting the photon at the point it hits the star. We
found that the fractional differences of the intensity and polarization values
(\di and \ddpol; see end of Section \ref{sec_output} for their definition)
which characterize a given model are less than 0.0044 and 0.023 respectively. 
These differences are small and within the errors of the simulations
(see Section \ref{sec_output}). All the results presented in this work,
consider the backwarming.

A spherically symmetric envelope produces a given total amount of
non-polarized flux independent of the observer's direction.  In contrast,
the flux and polarization of an envelope with one or more blobs depend
on the direction of the observer. Consequently, the photopolarimetric
variability may arise in two situations: (1) when the wind changes from
a homogeneous configuration to an inhomogeneous one; or (2) when the
wind is always inhomogeneous, but with blobs whose position relative
to the observer is variable.  The resultant variations from situation
(2)  may actually be observationally very similar to case (1). This is
because an envelope with a blob may be practically indistinguishable
from a homogeneous, one except if observed under some lines of sight
(see Section \ref{sec_output} and Figure \ref{fig-2d}). A moving blob may then
cause a change in the observed flux and polarization.

The input parameters of the models are listed in Table \ref{tbl-grid}
as well as the range of values for which we have obtained results. There
is a relatively large number of input parameters but, as we will see in
Section \ref{sec_wr}, we were able to constrain some of them based on the
observed variation of the intensity and polarization. The results do not
depend on the linear scale of the model. If we modify the source,
blob and envelope sizes using a same arbitrary multiplicative factor,
the output is exactly the same, provided the optical depth of the envelope
is kept the same. Thus all the parameters with units of length have been
normalized to the radius of the source. Because of that, the source
radius is not strictly an input parameter. All the models have used
the same source and external envelope radii (Table \ref{tbl-grid}). The
density of most models has been considered as constant. We have indicated
where we have used another density law.

The reliability of the results has been tested in different ways.
For instance, our code provides values of the polarization consistent
with zero in isotropic configurations and the results do not change
if the scale of the problem varies.  Other quantitative tests were
performed by comparing our results with those of previous calculations.
We have used for the latter: (1) the analytical solution of Chandrasekhar
(1950) for a semi-infinite plane-parallel optically thick envelope;
(2) the Monte Carlo solution of \cite{ld88} for a
dusty envelope; (3) the Monte Carlo solution of  Code \& Whitney (1995)
for a blob illuminated by a plane-parallel beam; (4) the analytical
results of Brown (1994). Our results are consistent with all the above
calculations. Graphs and tables can be found in \cite{rod97}.

\subsection{Model output}
\label{sec_output}

The code provides values of the flux, linear polarization and its position
angle, and the circular polarization as a function of the line of sight
under which the system is observed. These directions are represented by a
spherical coordinate system centered on the star: $\theta$ represents the
polar angle and $\phi$, the azimuthal angle. The photons are classified
in bins of same solid angle whose centers are equally spaced in
$\phi$ and $\cos\theta$ (it is necessary to use $\cos\theta$ for the
bins to have the same solid angle). The emerging circular polarization
is null in an electron scattering wind, if no circular polarization is input
at the base of the wind, since this process does not introduce any phase shift
upon the scattered wave. The flux is normalized to the value of a homogeneous
wind. Here, a value equal to one means that the flux is the same as that
observed from a wind without blobs.

Figure \ref{fig-2d} shows an example of the angular dependence of the
flux and polarization obtained with our code. The blob is located at
$\theta~=~90\arcdeg$ and $\phi~=~180\arcdeg$. The input parameters of this
model are given in the caption.  The flux as a function of the line of
sight direction, shown in Figure \ref{fig-2d}a, has a relatively simple
pattern. The flux has been normalized to that of a homogeneous envelope,
$I_o$, so $I/I_o$ is shown.  The central dip is caused by extinction by
the blob.  The light extinguished from along the direction of the blob
is scattered to all directions according to the phase function of the
scattering process which causes a flux modulation away from the line of sight
to the blob. In the figure, the maximum of the scattered flux around
90$\arcdeg$ from the blob direction can be noticed. This increase in flux is
much smaller than the decrease caused by extinction along the blob direction.

Figure \ref{fig-2d}b illustrates the polarization pattern which is
primarily determined by the electron scattering phase function. Therefore,
the polarization is approximately zero near the direction of the blob and
near the opposite direction ($\theta~=~90\arcdeg$,$\phi~=~0\arcdeg$). Its
maximum value is reached near 90$\arcdeg$ from the blob direction;
the actual angle depends on the optical depths of the envelope and blob.

The errors in our Monte Carlo simulations follow a Poisson statistics;
consequently they decrease as $\sqrt{N}$ when a larger number of photons
is used. For the number of photons and solid angle bins used, the uncertainty in
the normalized intensity values derived using the Poisson statistics is in the
third significative digit: $3.6 \times 10^{-3}$ for $3 \times 10^7$ photons
and $1.2 \times 10^{-3}$ for  $3 \times 10^8$ photons. These values agree
with the standard deviation of the intensity in regions where it is expected
to be constant.

In order to estimate the error in the polarization, we have run some
models for which we expect a null value of the polarization for any
direction. In this case, the average value of the polarization can be
assumed to give an estimate of the uncertainty. This value is practically
the same as the Poisson error derived from all the simulated photons:
$0.018\%$ for $3 \times 10^7$ photons and $0.0057\%$ for $3 \times 10^8$
photons.

An additional remark should be made about the polarization errors. Because the
polarization is an absolutely positive number, its error biases the
polarization towards larger values: the best estimate to the polarization is
given by $\sqrt{P_{obs}^2-\sigma_P^2}$, where $P_{obs}$ is the observed
polarization and $\sigma_P$ is the associated error (Clarke \&
Stewart 1986).  Because of that, the polarization degrees obtained in
the simulations may overestimate the actual values.

In order to simplify the analysis, each model was reduced to two values:
the maximum variation of the flux, $\Delta I$; and an estimate of the
maximum polarization, $\Delta P$, as described below.  These values
are the ones to be compared with the WR photometric and polarimetric
variations.  The value of $\Delta I$ is given by: $1 - (I_{min}/I_o)$,
where $I_{min}$ is the minimum value of the flux.  The value of
\di for the model in Figure \ref{fig-2d}a is 67\%. In doing that,
we assume that the flux variation in WR stars is produced by the passage of
a blob in the line of sight causing a \emph{decrease} in the total flux
from the star, i.e. the changes in flux are due to the extinction. As
illustrated by Figure  \ref{fig-2d}, the blobs also scatter, so the flux
increases along directions which do not intercept the blob.  However, as
we will further discuss towards the end of Section \ref{sec_dep} and in Section
\ref{sec_scat}, this variation is smaller than that caused by extinction if
$N_{bl}$ is less or around 20, for optically thin envelopes.

The polarization from a homogeneous envelope is zero in any direction.  The
polarization caused by an envelope with blobs can be then considered as the
polarimetric variation itself.  The maximum polarization was estimated in
the following way. We first calculated the average polarization for bins
of same $\theta$.  The value of $\Delta P$ was chosen as the maximum one
from such averages. This procedure enables us to have also an estimate of
the variance of the maximum polarization.  It also avoids any bias that
may result from selecting the largest absolute value among all the bins.
It could be argued that the bins of same $\cos\theta$ do not all form
the same angle with the blob, so that an average might not be appropriate.
However, given that the bin sizes are relatively large and that the
maximum polarization occurs at angles close to 90$\arcdeg$ from the blob,
this average is a valid approximation. The value of \dpol so calculated for
the model shown in Figure  \ref{fig-2d}b is $0.86\% \pm 0.08\%$. The 
maximum value of the polarization for the same model is 0.98\%.
 
\section{General results}
\label{sec_results}

In this section, we present some numerical results which are helpful in
order to understand the model physics.  We first present the dependence
of the results on the input parameters.  We then show specific results
contrasting the extinction and the scattered flux obtained with the
model. We do not contend that all models here presented are valid
representations of inhomogeneous WR envelopes, since some of them are
very optically thin.

\subsection{Dependence of the results on
the input parameters}
\label{sec_dep}

A general drawback of Monte Carlo simulations of physical problems
is that the influence of a given parameter may tend to be lost in the
final result.  To gain an insight into this influence, we present in this
section the behavior of \di and \dpol as function of some parameters of
the model.

\paragraph{Optical depth of the envelope}

The dependence of \di and \dpol with the optical depth of the envelope is
shown in Figure \ref{tau-env} (solid and dotted lines). The
polarimetric and photometric variations decrease as $\tau_{env}$ increases. 
This happens because the higher the envelope density, the higher the number of
scatterings in the envelope which tends to mask any effect produced by the
blob.

\paragraph{Optical depth of the blob}

Figure \ref{tau-env} also illustrates the effect of increasing the
optical depth of the blob. For very small values, when single scattering
dominates, \di and \dpol increase linearly with $\tau_{bl}$ (it cannot be
properly seen in the figure because of the abscissa scale). With further
increase in the blob optical depth, multiple scattering becomes more
important and \di and \dpol increase much more slowly and do not seem
to exceed a limiting value. The radiation can penetrate optically thin
blobs, so all the scatterers contribute to the variations, specially the
polarimetric ones. On the other hand, for optically thick blobs, only the
scatterers placed in the region facing the source play a role in the polarized
flux. Both \di and \dpol then level off with the increase in $\tau_{bl}$; \dpol
levels off at a smaller value of $\tau_{bl}$ compared to \di since the
polarization is more sensitive to the increase in optical depth.

\paragraph{Distance of the blob to the central source}

The solid angle subtended by the blob decreases when it is located
farther from the source. Therefore, a smaller fraction of the source
radiation is intercepted by the blob and \dpol decreases (Figure 
\ref{tau-env}; compare the solid and dashed curves). The average optical depth
of a bin, particularly that of the bin including the direction passing through
the blob center,  is smaller if the blob is located at a larger distance from
the source, so \di also decreases with distance (Figure \ref{tau-env}).
Note that the effect on \dpol is real and does not depend on the bin size,
while that on \di is caused by the finite spatial resolution of the models.

\paragraph{Blob size}

Figure \ref{fig-size} shows the behavior of \di and \dpol as a function of
blob size. It can be noticed that the variations depend strongly on the
blob dimension. This rapid increase of \di and \dpol with $R_{bl}$ is connected
with two important physical parameters: the solid angle of the blob and the
total number of scatterers in the blob. Each of these parameters dominates over
a certain range of the blob optical depth. In the optically thin limit, all
particles in the blob effectively scatter the light. However, for optically
thick blobs, only the particles near the source are important in the
scattering process. In larger blobs the area facing the source is larger, so
larger is the overall variation in intensity and polarization of the emerging
flux. Comparing two models of same blob optical depth, $\tau_{bl}$, but
different sizes, $R_{bl}$, we note that the blob having a larger size has a
smaller density in order to keep the same optical depth.

\paragraph{Different density laws}

In Figure \ref{fig-lei}, we compare the results for two different
density laws: uniform and beta laws. We have adopted a beta coefficient
of 0.5 in these runs. This is the value obtained for the most
simple calculations of a radiative wind (Castor, Abbott \& Klein 1975). We
recall that the blob follows the same density law as the envelope.  The
dependence of \di and \dpol with $\tau_{bl}$ are shown for the optically thin
($\tau_{env}$~=~0.01) and thick ($\tau_{env}$~=~1) cases. In the latter, in
spite of the differences in the distribution of matter along the envelope, the
variations in the flux and polarization of the emerging light are not
considerably affected by the density radial distribution. For the
optically thin cases, the differences are larger, particularly in the
intensity curves. The optical depth of the model envelope is very small
($\tau_{env}$~=~0.01) and the difference cannot be due to scattering in
the envelope. When the matter follows a beta law it is more
concentrated near the base. For radial optical paths passing
through the center of the blob, the optical depth is the same for the
two density laws - this is imposed by the choice of parameters.  Other
radial paths, however, have smaller optical depth in the case of the
beta law.  As our solid angle bins have a finite size, they include a
range of directions and the {\it effective} optical depth in the bin is
smaller in the beta law case. This causes the smaller \di for the beta
law shown in Figure \ref{fig-lei}.

The differences in \di and \dpol between the two density laws are small
considering our relatively  simple model and that we do not have very
precise observational values to be compared with. For these reasons we
consider that uniform envelopes give a sufficient representation for
scattering calculations in the present context.  This result is
important because the beta law calculations are computationally (much)
more costly than uniform density ones.

\paragraph{Number of blobs}

The presence of a number of sub-peaks in emission lines is usually
taken as evidence for the presence of more than one blob in WR winds
(Robert 1994). The average number of such sub-peaks is around 10 (Robert 1994).
It is thus important to understand how \di and \dpol can be affected by the
number of blobs as compared to the values produced by one condensation. This
section does not intend to fully explore the possibilities of an optically
thick envelope with many blobs; this merits a separate paper and will be done
elsewhere. Single scattering calculations for a cloud of point-like blobs have
been presented by Richardson et al. (1996).

We have run some models with a number of blobs, $N_{bl}$, between 1 and
75, assuming no superposition among blobs. In Table \ref{tbl-number} we
summarize the results. All the parameters where considered fixed (see Table
\ref{tbl-number} for their values) across simulations, except for $N_{bl}$ and
the blob positions. The exact meaning of the columns is explained in the
following paragraphs. Generally speaking, they are averages of a number of
simulations (between 3 and 10, see Table \ref{tbl-number}) with different
spatial blob distribution. We have assumed that the blobs have a uniform
angular distribution, but with a fixed distance to the source ($d_{bl}~=~3$).
The $\sigma$ values are the standard deviations of the sample. 

Table \ref{tbl-number} shows the average of \di for sets of simulations with
the same $N_{bl}$. An increase of \di with the number of blobs is apparent. As
the spatial resolution of the models is modest, the position of the blob
relative to the bin center could have some effect on \ddi. In order to
ascertain  such effect, we ran 10 simulations with a single blob; in each of
them the blob was located at different positions in the envelope. The first
line in Table \ref{tbl-number} is the average of these simulations. The
resultant small error means that \di is quite  independent of the blob
position. Thus the increase in \di with $N_{bl}$ is not caused by effects
related to the blob position, but it seems to be a consequence of the presence
of more than one blob in the envelope. The explanation for this effect is the
following. Let's consider only photons emitted by the source in one direction.
They form a beam whose cross section has the size of the stellar disk ($\pi
R_{s}^2$). In the simulations we have run, the cross section of one blob is
0.25 of this value. If the envelope has only one blob, obviously the beam can
only cross one blob. But with the increase of $N_{bl}$, more than one blob can
be crossed by the same beam: it will depend on the angular distribution of the
condensations in each simulation. This explains the increase of \di with the
number of blobs and also the larger standard deviation compared with the case
of one blob. Table \ref{tbl-number} also shows that \di increases until
$N_{bl}$ around 30 and then levels off or decreases. This happens because the
scattered photons becomes an important fraction of photons leaving the
envelope and, as their angular distribution tends to be isotropic, this 
"emission" tends to fill the dips in the intensity caused by extinction. This
effect has the same nature of the decrease of \di when $\tau_{env}$ increases
(see Figure \ref{tau-env}).

In Table \ref{tbl-number} we also list the polarization as a
function of $N_{bl}$. The fourth column shows the averages of the
polarization degree, $\bar{P}$, from a given set of simulations. The fifth
column presents the standard deviation of the sample, $\sigma_{P}$; this
gives an idea of how wide the distribution of polarization actually is. 
Figure \ref{fig-hist} shows the histograms  of the polarization degree for
simulations of same $N_{bl}$. The values in  Table \ref{tbl-number} are then
a summary of these histograms. The maximum of $\bar{P}$ occurs around 20 blobs
and after that $\bar{P}$ decreases slowly. Around this number, the standard
deviation of the polarization also peaks. This may be interpreted as the
situation where the anisotropy of the envelope is maximum; this is valid for
the configuration represented by the model parameters: blob size, optical
depths of blob and envelope, blob distance to the source, and so on. These
results may be contrasted with those of Richardson et al. (1996): for
distinct model assumptions, the average polarization and variance were always
increasing functions of the blob number. The largest number of blobs they have
considered was 100.

\subsection{Extinction versus the scattered flux}
\label{sec_scat}

In Table \ref{tbl-scat-ext}, we show some results related to the scattered
flux of our models. This table also presents the variation caused by
extinction to comparison. In the first three lines, we show the results for
one-blob models whose the variation in flux due to scattering, $F_{scat}$,
was the highest for a given optical depth of the envelope. $F_{scat}$
was calculated considering the largest flux among all the bins in one given
simulation. In the remaining lines of this table, we fixed all the parameters
of the models, except $N_{bl}$. The values of $F_{scat}$ are the averages of
different simulations with the same $N_{bl}$.

If we consider only the results for one-blob envelopes, we notice that:
(1) $F_{scat}$ is less than 5\% for the optically thicker envelopes; (2) the variation caused by
extinction is always greater than that caused by scattering. On the other
hand, for envelopes with a large number of blobs ($N_{bl}$ $>$
$20$), the amount of scattered flux can be very similar to the extinguished
one. The value of $N_{bl}$ for which $F_{scat}$  becomes important is
dependent on the blob parameters (size and optical depth), since
it is a main function of the total number of scatterers. Our results 
for many blobs were obtained for optically thin envelopes. The one-blob
results (Table \ref{tbl-scat-ext}) seem to suggest that $F_{scat}$
will be less important for optically thick envelopes in the many-blob
situation as well. This is quite plausible since in optically thick envelopes
the flux tends to be more isotropic.

\section{Random broad band variability in WR stars}

\subsection{Simulation results}
\label{sec_wr}

Single WR stars may present polarimetric and flux variations of 0.5\%
and 10\%, respectively. From the previous sections, we can state that the
variations observed in flux can be produced by extinction, if $N_{bl}$ is
small, or by extinction and scattering, if $N_{bl}$ is larger than around 10,
for the parameters used.  If the brightness changes are caused by
extinction, the polarization will never be observed simultaneously with the
variation in intensity produced by the same blob. If, for instance, a given
blob is along the line of sight towards the center of the star it will cause a
decrease in flux, but the (maximum) polarization caused by the same blob will
only be noted by an observer in the direction perpendicular to that one. In
other words, the largest variations in flux and polarization do not occur
along the same line of sight. So it would not in principle be necessary to
consider only one type of blob to explain the observations. We nevertheless
may assume here without loss of generality that the observed fluctuations are
produced by blobs having the same physical characteristics. As we now show,
this assumption explains the observed photopolarimetric fluctuations.

Some results for one-blob models with $\tau_{env}$ greater or equal to 0.1 are
shown in Table \ref{tbl-models}. They were selected from hundreds of
simulations whose parameters are within the ranges presented in Table
\ref{tbl-grid}. We have selected models with $\Delta I$ around 10\% and
$\Delta P$ greater than 0.2\%.  Some models have optically thin
envelopes which may not correctly represent a WR envelope.  However we
have kept them to allow a comparison with the optically thick model
results. In Table \ref{tbl-models} we also include the mass of one
blob, $M_{bl}$, as a fraction of the total envelope mass for each
model. These values rely on the composition and the degree of
ionization not varying strongly throughout the envelope since they
depend on the electron density.

A general result is that an extinction value of 10\% is easily achieved
for practically any choice of blob and envelope parameters. One needs
only to adjust the optical depth of the blob, $\tau_{bl}$. Although it
is not shown in Table \ref{tbl-models} (because of the associated small
values of polarization), models with blobs far from the base of the
envelope can also produce 10\% of variation in flux.  We conclude that
the photometric variation by itself does not strongly constrain the physical
characteristics of the blobs.

On the other hand, a polarization of $\approx$ 0.5\% is only obtained
for a few specific cases. In general, a blob produces smaller values of
polarization. The highest values of polarization are achieved for an
envelope with a large blob near the base of the envelope: for instance, the
model with $\tau_{env}$~=~1.0, $d_{bl}$~=~3, $R_{bl}$~=~1.0 and a beta law for
the density is one such case. We conclude that the blobs must have a
large geometric cross section in order to produce the observed values
of polarization (see also Section \ref{sec_large_cs}).

We can now address the applicability of these results to the flux
variations of WR stars. The simplest scenario is that of an envelope
with a single blob.  In this situation, we have already shown that the
scattered flux does not produce a variation as high as 10\%, while the
extinction can reach this value easily. The question is, how probable
is it to observe such extinction? The exact answer depends on the size,
distance to the source, and density distribution of the blob, but we
can notice in Figure  \ref{fig-2d} that the range of directions where we
can observe the extinction is considerably large. So the probability of
obtaining around 10\% from extinction from the blob is non negligible. With an
increase in the number of blobs, this probability increases because there will
be more lines of sight which intercept a blob, whereas the angular shape and
level of the decrease caused by extinction remains practically the same: this
is valid if the blobs are not very close angularly. However, the scattered
flux also increases with the number of blobs because each blob contributes
individually to the total (see Table \ref{tbl-scat-ext}). We note however
that the results for many blobs (end of Section \ref{sec_dep} and Section
\ref{sec_scat}) were obtained for optically thin envelopes: this may
overestimate $F_{scat}$. Therefore, in a more realistic scenario where many
blobs coexist, the flux variations will originate from the interplay between
two situations: (1) the line of sight intercepts any blob, in which case
extinction effects dominate; (2) there is no blob in the line of sight, in
which case there is a scattered light contribution dependent of the number of
blobs and their spatial distribution.  This result contrasts with previous
studies (Brown et al. 1995; Richardson et al.  1996), where it was assumed
that the flux variation is produced mainly by scattering.

The polarization produced by an inhomogeneous envelope depends in a more
complicated way on the envelope parameters than the flux. The main reasons
are: (1) polarizations from individual blobs do not sum algebraically; (2) a
blob produces polarization in regions not confined to a line of sight. If we
consider only one blob, we have shown that a polarization of 0.5\% is only
achieved with blobs near (less than about 3 stellar radius) the base of the
envelope and whose sizes are similar to the stellar radius. Figure
\ref{fig-hist} shows some examples of the number distribution of polarization
degree for simulations with a varying number of blobs for optically thin
envelopes. Diagrams like those can be compared with observational data in order
to constrain the physical characteristics of the blob. They would also be
helpful in predicting the polarization from a given theoretical blob model.
This will be explored in a forthcoming paper.

We have not included the intrinsic emission from the envelope and blobs.
It would probably tend to smooth the variations in flux and polarization
with aspect, since the emission is isotropic and non-polarized.

As stressed by Brown et al. (1995), simultaneous observations in the
three mode of observations would be very important for further
constraining the blob properties.  We are carrying an observational
program to observe highly variable WR stars using photopolarimetric and
spectroscopic techniques.

\subsection{Analytical calculations for the blob cross section}
\label{sec_large_cs}

A simple argument in favor of a large cross section for a blob in order
to cause the observed values of \di and \dpol can be presented. Let us
assume the situation of a blob immersed in a transparent envelope, with
the blob and the star in the plane of sky so the scattered flux from
the blob has a 90$\arcdeg$ scattering angle. The unscattered brightness
(light coming directly from the source) received by the observer will
be called $I_{dir}$. We assumed that this component is completely
unpolarized. 

Assuming that the blob scatters the light it intercepts according to
the phase function of the electron scattering, the scattered intensity
from a blob to the observer direction, $I_{scat}$, is:

\begin{equation}
I_{scat}=\frac{3}{8} \frac{1}{2\pi} I_{dir} \sigma_{bl},
\end{equation}

\noindent where $\sigma_{bl}$ is the solid angle of the blob (as seen by
the central source) and the numerical factor before $I_{dir}$ is the
probability of scattering in the observer direction. The scattered light is
100\% polarized, so the net polarization is just the ratio between $I_{scat}$
and ($I_{scat}$ + $I_{dir}$) (Code \& Whitney 1995).

In this simple scenario, a blob of one stellar radius situated at 2 stellar
radius produces a polarization of 4\%. If the blob has half a stellar radius,
the polarization will be 1\%. These values are an overestimate for the
polarization:  if the blob's optical depth is small, only a fraction of
the source light will be scattered; consequently $I_{scat}$ will
decrease and so will the polarization. On the other extreme, if the
blob is optically thick, the scattered light, $I_{scat}$, would be much
more isotropic and not 100\% polarized because of multiple scattering inside
the blob. In any case, in order to achieve a degree of polarization around
0.5\% the blobs must have large cross sections.

\section{Photopolarimetric variability in other early-type stars}

Taylor et al. (1991) have shown that P Cygni, a luminous blue variable,
presents random variations in its optical polarization.  Their data,
collected during 1989 and 1990, show amplitude variations between 0.04\%
and 0.48\%, with no preferred value for the position angle. They have
pointed out that such a large variation indicates that the material
causing polarization must be close to the star and suggested that a
clumpy and variable wind must be causing the variation in continuous
polarization. Coronographic images of this star shows indeed unresolved
clumps of emission distributed asymmetrically in the envelope (Nota et
al. 1995). Consequently, the model of an envelope with condensations is
consistent with P Cygni observations.

Polarimetric observations of OB supergiants have shown that a large
fraction of these stars (7 out of 10) presents random fluctuation between
0.2\% and 0.4\% (Lupie \& Nordsieck 1987; Bjorkman 1994). The existence
of blobs in the wind which scatter the source light has been considered as a
possible explanation to this behavior.

It is clear therefore that random variability is also present in other
early-type stars besides WR stars. An envelope with inhomogeneities is
a possible interpretation in these cases and seems indeed to be the best
explanation, at least for the supergiant P Cyg. Based on the observed
polarization variations, the blobs seem to have properties similar to those which
explain the WR variability: they must be large and close
to the star.  This may be an indication that the same phenomenon is working
in hot stars with quite distinct winds and causing a similar clumpy
structure.

\section{Conclusions}

We have for the first time treated the problem of the multiple scattering
in an electron scattering envelope where regions of enhanced density
exist. The model results, obtained using a Monte Carlo code, were used
to study the random variation in broad band flux and polarization of
WR stars.  We show that the flux variation amplitude may be due to the
extinction caused by blobs intercepting the line of sight, if $N_{bl}$ is
smaller than 10. For an envelope with a larger number of blobs, both
scattering and extinction, may play a role in the flux variations.  We show
further that the flux variations do not constrain much the physical properties
of the inhomogeneities. The polarization variation, on the other hand, used
simultaneously with the photometric variation, is a more useful constraint to
the blob characteristics. The photopolarimetric variability in WR stars
indicates that the inhomogeneities must have relatively large sizes
(comparable to that of the star) and be near the base of the photosphere.  The
envelope may possibly, but not necessarily, have few blobs.  If a clumpy
structure is causing the polarimetric fluctuations seen in other types of hot
stars, the blobs probably have similar characteristics since the amplitude
variations are of the same order.

\acknowledgments

We thank J. Bjorkman for his careful reading of the paper and his valuable
comments. We also thank the anonymous referee for his insightful comments and
remarks which helped to improved the paper. We are also grateful to Fapesp
for the financial support to the polarimetric group at IAG/USP (grants
92/3345-0, 94/0033-3 and 97/11299-2).  CVR also thanks Fapesp by its financial
support through grants 1992/1812-0 and 1998/1443-1. AMM additionally
acknowledges Fapesp grant 98/4267-0 and CNPq grant 301558/79-5. This research
has made use of NASA's Astrophysics Data System Abstract Service.

\clearpage

%
%

%
\clearpage
%



\begin{figure}
\caption{Example of the angular dependence of intensity and polarization
obtained as result of a Monte Carlo simulation for the radiative
transfer in an inhomogeneous envelope with a single blob of radius $R_{bl}$, located at
$\theta~=~90\arcdeg$ and $\phi~=~180\arcdeg$: (a) intensity relative to that
of the homogeneous case; (b) percent polarization. The input parameters
are: $\tau_{env}$~=~0.01; $d_{bl}$~=~3; $R_{bl}$~=~1; $\tau_{bl}$~=~10;
uniform density law. The envelope radius, $R_{env}$, has been taken as 10 in
this and all other simulations in this paper. $R_{env}$, $R_{bl}$, $d_{bl}$
are normalized to the source radius.}
\label{fig-2d}
\end{figure}

\begin{figure}
\caption{Variations of intensity and polarization in one-blob envelopes
of different optical depths and at different distances from the
source. The points represented by triangles are upper limits to the
polarization. The model parameters are:  $R_{bl}$~=~1; uniform
density law.}
\label{tau-env}
\end{figure}

\begin{figure}
\caption{Variations of intensity and polarization caused by one blob
of variable size. The envelope density follows an uniform law.}
\label{fig-size}
\end{figure}

\begin{figure}
\caption{Variations of intensity and polarization caused by models with
different density laws with a single blob. The optical depths of the
envelope are 0.01 and 1 for the optically thin and thick cases,
respectively.  The model parameters are: $d_{bl}$~=~3; $R_{bl}$~=~1.0.
The beta law coefficient used was 0.5.}
\label{fig-lei}
\end{figure}

\begin{figure}
\caption{Histograms of the number frequency of the polarization degree for
simulations with varying $N_{bl}$ (1 through 75). The models parameters are:
$\tau_{env}$~=~0.01; $\tau_{bl}$~=~5.0; $R_{bl}$~=~0.5; $d_{bl}$~=~3.0.}
\label{fig-hist}
\end{figure}

\clearpage
%
%
%
\begin{deluxetable}{ll}
\tablecaption{Input parameters used to generate models
of inhomogeneous WR winds
\label{tbl-grid}}
\tablewidth{0pt}
\tablehead{
\colhead{Input Parameters} &
\colhead{Range of Values}
}
\startdata
Source radius, $R_s$ & 1.0\\
Envelope radius, $R_{env}$\tablenotemark{a} & 10 \\
Envelope optical depth, $\tau_{env}$ & 0.01 - 2 \\
Blob radius, $R_{bl}$\tablenotemark{a} & 0.25, 0.5, 1.0 \\
Blob optical depth, $\tau_{bl}$ & 0.1, 0.2, 0.5, 1.0, 2.0, 5.0, 10.0 \\
Blob distance, $d_{bl}$\tablenotemark{a} & 3, 5, 7, 9 \\
Number of blobs, $N_{bl}$  & 1 - 75 \\
\enddata
\tablenotetext{a}{Normalized to $R_s$.}
\end{deluxetable}
%

%
%
\clearpage

\begin{deluxetable}{cccccc}
\tablecaption{Summary of the results for models with a number of blobs
\tablenotemark{a} \label{tbl-number}}  
\tablewidth{0pt}
\tablehead{
\colhead{$N_{bl}$} &
\colhead{$\Delta I$} &
\colhead{$\sigma_{\Delta I}$} &
\colhead{$\bar{P}$} &
\colhead{$\sigma_{P}$} &
Number of \\
 &  
\colhead{(\%)} &
\colhead{(\%)} &
\colhead{(\%)} &
\colhead{(\%)} &
Simulations }
\startdata
 1 & 16.5 & 0.4 & 0.17 & 0.09 & 10 \\
 3 & 18.3 & 4.4 & 0.28 & 0.13 & 10 \\
10 & 26.0 & 4.5 & 0.39 & 0.18 & 10 \\
20 & 28.2 & 2.8 & 0.49 & 0.24 & 8 \\
30 & 27.3 & 3.0 & 0.44 & 0.23 & 3 \\
40 & 26.6 & 1.7 & 0.41 & 0.17 & 3 \\
50 & 26.4 & 3.4 & 0.42 & 0.16 & 3 \\
60 & 22.1 & 2.0 & 0.39 & 0.14 & 3 \\
70 & 22.3 & 2.0 & 0.40 & 0.14 & 4 \\
75 & 18.1 & 1.7 & 0.39 & 0.14 & 4 \\
\enddata
\tablenotetext{a}{The model parameters are: $\tau_{env}$~=~0.01;
$\tau_{bl}$~=~5.0; $R_{bl}$~=~0.5; $d_{bl}$~=~3.0.} 
\end{deluxetable}

\clearpage

%
%
\begin{deluxetable}{lllllll}
\tablecaption{Variation in flux caused by scattering and
extinction in an inhomogeneous WR envelope 
\label{tbl-scat-ext}}
\tablewidth{0pt}
\tablehead{
\colhead{$N_{bl}$} &
\colhead{$\tau_{env}$} &
\colhead{$\tau_{bl}$} &
\colhead{$R_{bl}$} &
\colhead{$d_{bl}$} &
\colhead{$F_{scat}$} &
\colhead{\di}\\
& & & & & \colhead{\%} & \colhead{\%}
}
\startdata
1 & 0.01 & 10. & 1. & 3. & 7.1 & 76 \\ 
1 & 0.40 & 10. & 1. & 3. & 4.6 & 47 \\ 
1 & 1.00 & 10. & 1. & 3. & 2.7 & 27 \\ 
\\
1 & 0.01 & 5. & 0.5 & 3. & 1.9 & 16.5 \\
3 & 0.01 & 5. & 0.5 & 3. & 3.4 & 18.3 \\
10 & 0.01 & 5. & 0.5 & 3. & 8.5 & 26.0 \\
20 & 0.01 & 5. & 0.5 & 3. & 15.7 & 28.2 \\
30 & 0.01 & 5. & 0.5 & 3. & 21.9 & 27.3 \\
40 & 0.01 & 5. & 0.5 & 3. & 24.4 & 26.6 \\
50 & 0.01 & 5. & 0.5 & 3. & 26.3 & 26.4 \\
60 & 0.01 & 5. & 0.5 & 3. & 25.3 & 22.1 \\
70 & 0.01 & 5. & 0.5 & 3. & 23.3 & 22.3 \\
75 & 0.01 & 5. & 0.5 & 3. & 25.4 & 18.1 \\
\\
\enddata
\end{deluxetable}

%
%
\clearpage

\begin{deluxetable}{ccccccccccc}
\tablecaption{Best models of inhomogeneous WR winds\tablenotemark{a} \label{tbl-models}}
\tablewidth{0pt}
\tablehead{
\colhead{$\tau_{env}$} &
\colhead{$d_{bl}$} &
\colhead{$R_{bl}$} &
\colhead{$\tau_{bl}$} &
\colhead{Density} &
\colhead{$\Delta I$} &
\colhead{$\Delta P$} &
\colhead{$\sigma_{\Delta P}$} &
\colhead{$\Delta I/\Delta P$} &
\colhead{$ M_{bl}$\tablenotemark{b}} \\
& & & & Law & 
\colhead{(\%)} &
\colhead{(\%)} &
\colhead{(\%)}}
\startdata
0.10 & 3.0 & 0.5 & 2.0 & Unif. & 11.3 & 0.266 & 0.047 & 42 & 0.023\\ 
0.20 & 3.0 & 0.5 & 2.0 & Unif. & 10.0 & 0.249 & 0.066 & 40 & 0.011 \\ 
0.40 & 5.0 & 1.0 & 2.0 & Unif. & 28.8 & 0.335 & 0.071 & 86 & 0.023 \\ 
0.40 & 3.0 & 1.0 & 1.0 & Unif. & 20.4 & 0.446 & 0.079 & 46 & 0.011 \\ 
0.50 & 3.0 & 1.0 & 1.0 &  $\beta$ = 0.5 & 9.7 & 0.36 & 0.11 & 27 &
0.016 \\ 
1.00 & 3.0 & 1.0 & 5.0 & Unif. & 24.8 & 0.514 & 0.062 & 48 & 0.023 \\ 
1.00 & 3.0 & 1.0 & 2.0 & $\beta$ = 0.5 & 12.5 & 0.47 & 0.15 &
27 & 0.016 \\ 
1.00 & 5.0 & 1.0 & 10.0 & Unif. & 23.5 & 0.364 & 0.070 & 65 & 0.045 \\ 
2.0 & 3.0 & 1.0 & 5.0 & Unif. & 10.6 & 0.548 & 0.057 & 19 & 0.011 \\ 
\enddata

\tablenotetext{a}{All models listed here are for one blob ($N_{bl}$~=~1).}
\tablenotetext{b}{Mass relative to the total mass of the envelope.}
\end{deluxetable}


\begin{thebibliography}{}
%
\bibitem[Antokhin et al. 1995]{ant95} Antokhin, I., Bertrand, J.~F.,
Lamontagne, R., Moffat, A.~F.~J., \& Matthews, J. 1995, \aj, 109, 817
%
\bibitem[ ]{bj94} Bjorkman, K.~S. 1994, Ap\&SS, 221,335
%
\bibitem[ ]{b94} Brown, J.~C. 1994, \apss, 221, 357
%
\bibitem[Brown et al. (1995)]{bro95} Brown, J.~C., Richardson, L.~L., Antokhin, I.,
Robert, C., Moffat, A.~F.~J., \& St-Louis, N. 1995, \aap, 295, 725
%
\bibitem[ ]{c50} Chandrasekhar, S. 1950, Radiative Transfer
(Oxford: University Press)
%
\bibitem[ ]{cak} Castor, J.~I., Abbott, D.~C. \& Klein, R.~I. 1975, \apj, 195,
157
%
\bibitem[ ]{cs86} Clarke, D., \& Stewart, B.~G. 1986, Vistas in Astronomy,
29, 27 %
\bibitem[Code \& Whitney (1995)]{cw95} Code, A.~D., \& Whitney, B.~A. 1995, \apj, 441, 400
%
\bibitem[ ]{d87} Drissen, L., St.-Louis, N., Moffat, A.~F.~J., \&
Bastien, P. 1987, \apj, 322, 888
%
\bibitem[ ]{fh94} Fox, G.~K., \& Henrichs, H.~F. 1994, \mnras, 266, 945
%
\bibitem[ ]{g095} Gayley, K.~G., \& Owocki, S.~P. 1995, \apj, 446, 801
%
\bibitem[ ]{hk98} Hamann, W.-R., \& Koesterke, L. 1998, \aap, 335, 1003
%
\bibitem[ ]{h84} Hillier, D.~J. 1984, \apj, 280, 744
%
\bibitem[ ]{kh94} Kaper, L., \& Henrichs, H.~F. 1994, \apss, 221, 115
%
\bibitem[Lefevre \& Daniel (1988)]{ld88} Lefevre, J., \& Daniel,
J.~Y. 1988, in Polarized Radiation of Circunstellar Origin, ed.
G.~V. Coyne, S.~J. et al.  (Vatican City State: Vatican Observatory), 523
%
\bibitem[]{lm96} Lepine, S. \& Moffat, A.~F.~J. 1996, ApJ, 466, 392
%
\bibitem[]{lm99} Lepine, S. \& Moffat, A.~F.~J. 1999, ApJ, 514, 909
%
\bibitem[]{lem99} Lepine, S., Eversberg, T. \& Moffat, A.~F.~J. 1999, AJ, 117,
1441 
%
\bibitem[]{ln87} Lupie, O.~L., \& Nordsieck, K.~H. 1987, \aj, 92, 214
%
\bibitem[]{mr99} Magalh\~aes, A.~M., \& Rodrigues, C.~V. 1999, in IAU Coll.
169: Variable and Non-spherical Stellar Winds in Luminous Hot
Stars, ed. B. Wolf, O. Stahl \&  A. W. Fullerton. (Heidelberg:
Springer), 49
%
\bibitem[ ]{m98} Marchenko, S.~V., Moffat, A.~F.~J., Eversberg, T.,
Hill, G.~M., Tovmassian, G.~H., Morel, T., \& Seggewiss, W. 1998, \mnras,
294, 642
%
\bibitem[Moffat \& Robert 1991]{mr91} Moffat, A.~F.~J., \& Robert, C. 1991,
in IAU Symp. 143: Wolf-Rayet Stars  and Interrelations with Massive Stars,
ed. K.~A. van der Hucht \& B. Hidayat (Dordrecht: Kluwer), 109
%
\bibitem[ ]{mr92} Moffat, A.~F.~J., \& Robert, C. 1992, in ASP Conf. Ser. 22:
Nonisotropic and Variable Outflows from Stars, ed. L. Drissen,
C. Leitherer \& A. Nota (San Francisco: Astron. Soc. of Pacific), 203
%
\bibitem[ ]{m94} Moffat, A.~F.~J., Owocki, S.~P., Fullerton, A.~W., \&
St.-Louis, N. (ed.) 1994, Instability and Variability  of Hot-star
Winds, \apss, 221
%
\bibitem[ ]{n95} Nota, A., Livio, M., Clampin, M., \& Schulte-Ladbeck, R.
1995, \apj, 448, 788
%
\bibitem[ ]{o94} Owocki, S.~P. 1994, \apss, 221, 3
%
\bibitem[ ]{ps92} Prinja, R.~K., \& Smith, L.~J. 1992, \aap, 266, 377
%
\bibitem[Richardson et al. 1996]{r96} Richardson, L.~L., Brown, J.~C., \& Simmons, J.~F.~L. 1996, \aap, 306, 519
%
\bibitem[ ]{r94} Robert, C. 1994, \apss, 221, 137
%
\bibitem[ ]{r89} Robert, C., Moffat, A.~F.~J., Bastien, P., Drissen, L., \&
St.-Louis, N.  1989, \apj, 347, 1034 \
%
\bibitem[Rodrigues (1997)]{rod97}Rodrigues, C.~V. 1997, PhD Thesis,
Inst. Astron\^omico e Geof\'\i sico, Univ. de S\~ao Paulo
%
\bibitem[ ]{rm94} Rodrigues, C.~V., \& Magalh\~aes, A.~M. 1994, in
IAU Symp. 163: Wolf-Rayet Stars: Binaries, Colliding Winds and
Evolution, ed.  K.~A. van der Hucht \& P.~M. Williams (Dordrecht:
Kluwer), 260
%
\bibitem[ ]{sl87} St.-Louis, N., Drissen, L., Moffat, A.~F.~J., Bastien, P., \& Tapia, S. 1987, \apj, 322, 870
%
\bibitem[ ]{t91} Taylor, M., Nordsieck, K.~H., Schulte-Ladbeck, R.~E., \&
Bjorkman, K.~S. 1991, \aj, 102, 1197
%
\bibitem[ ]{ws99} Wolf, B., Stahl, O. \& Fullerton, A.~W. (ed.) 1999,
IAU Coll.  169: Variable and Non-spherical Stellar Winds in
Luminous Hot Stars (Heidelberg: Springer), in press
%
\end{thebibliography}
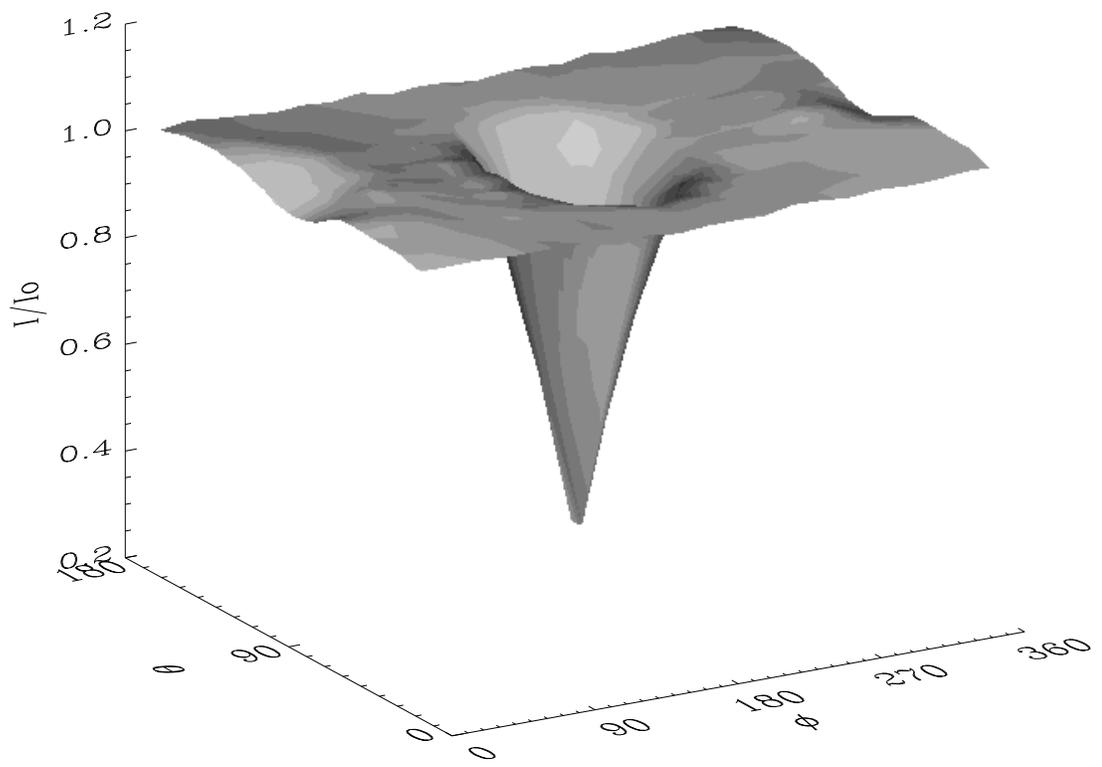
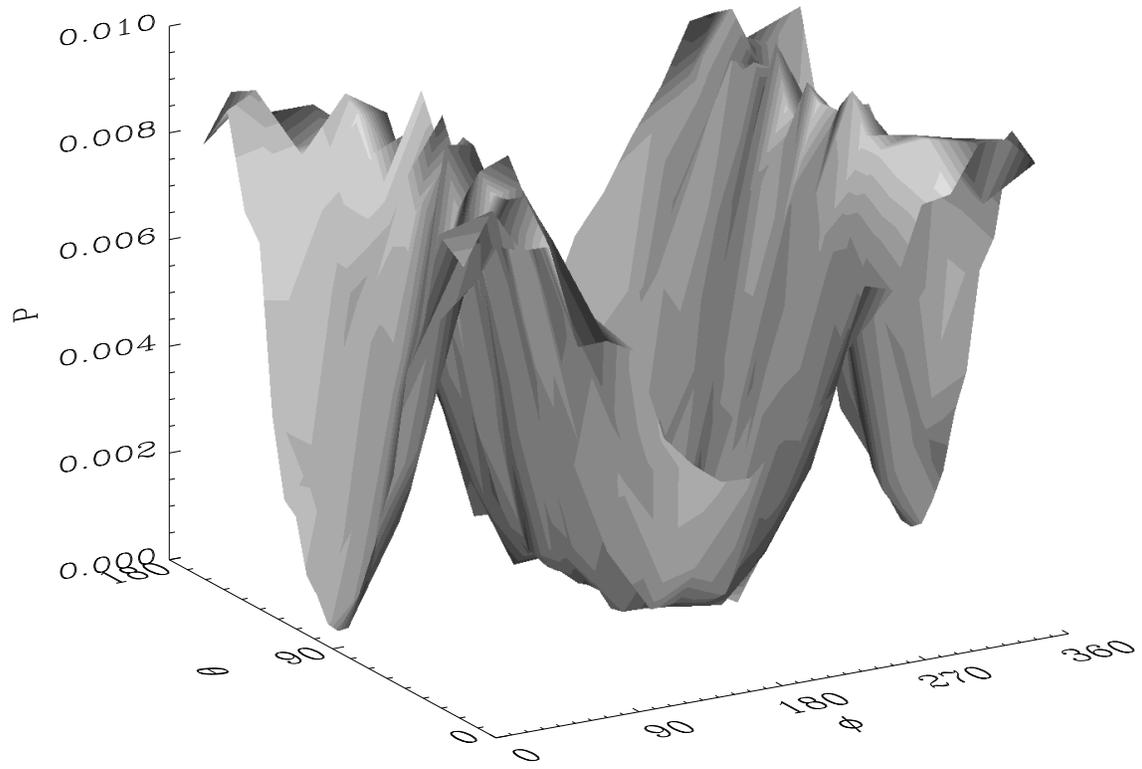
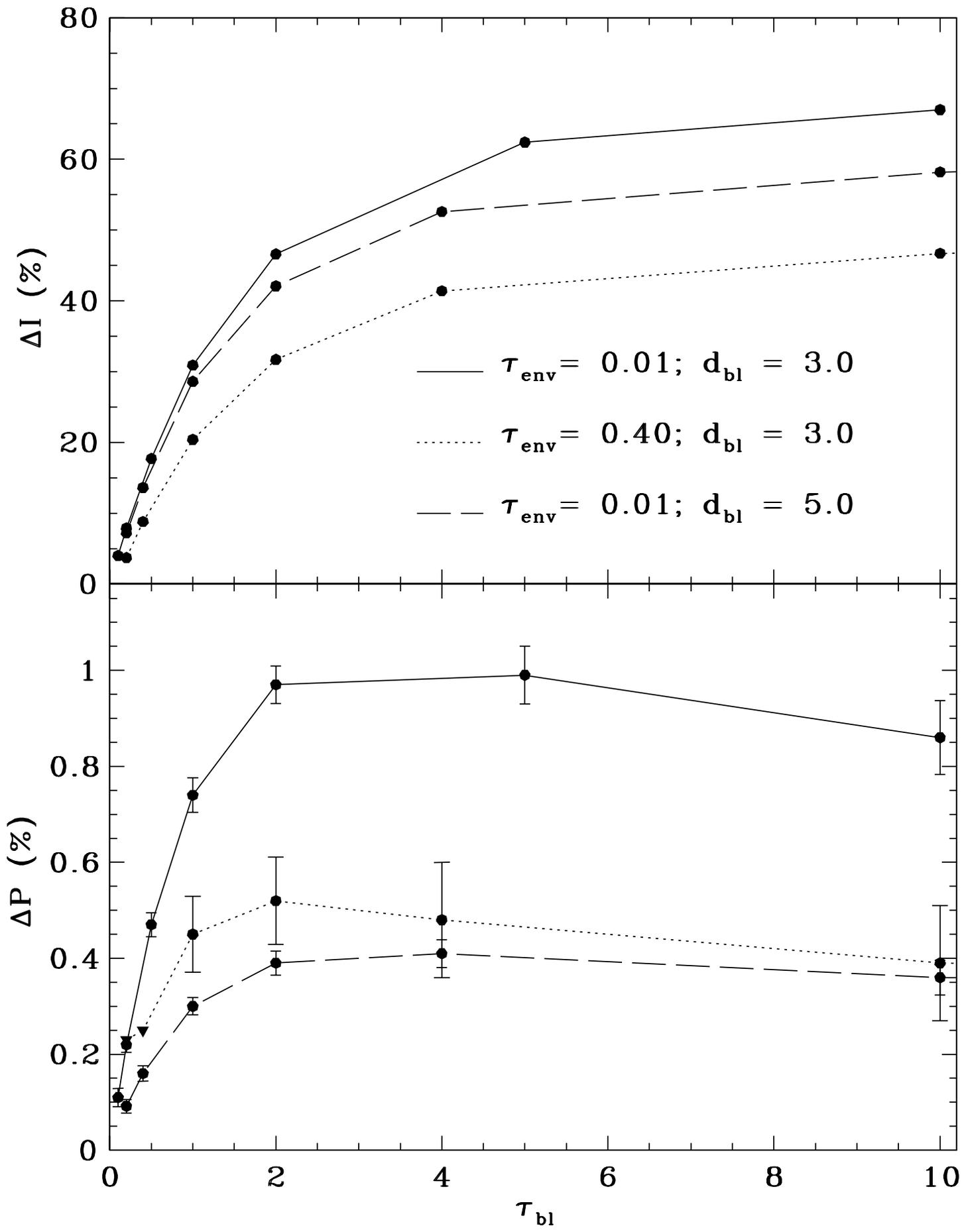
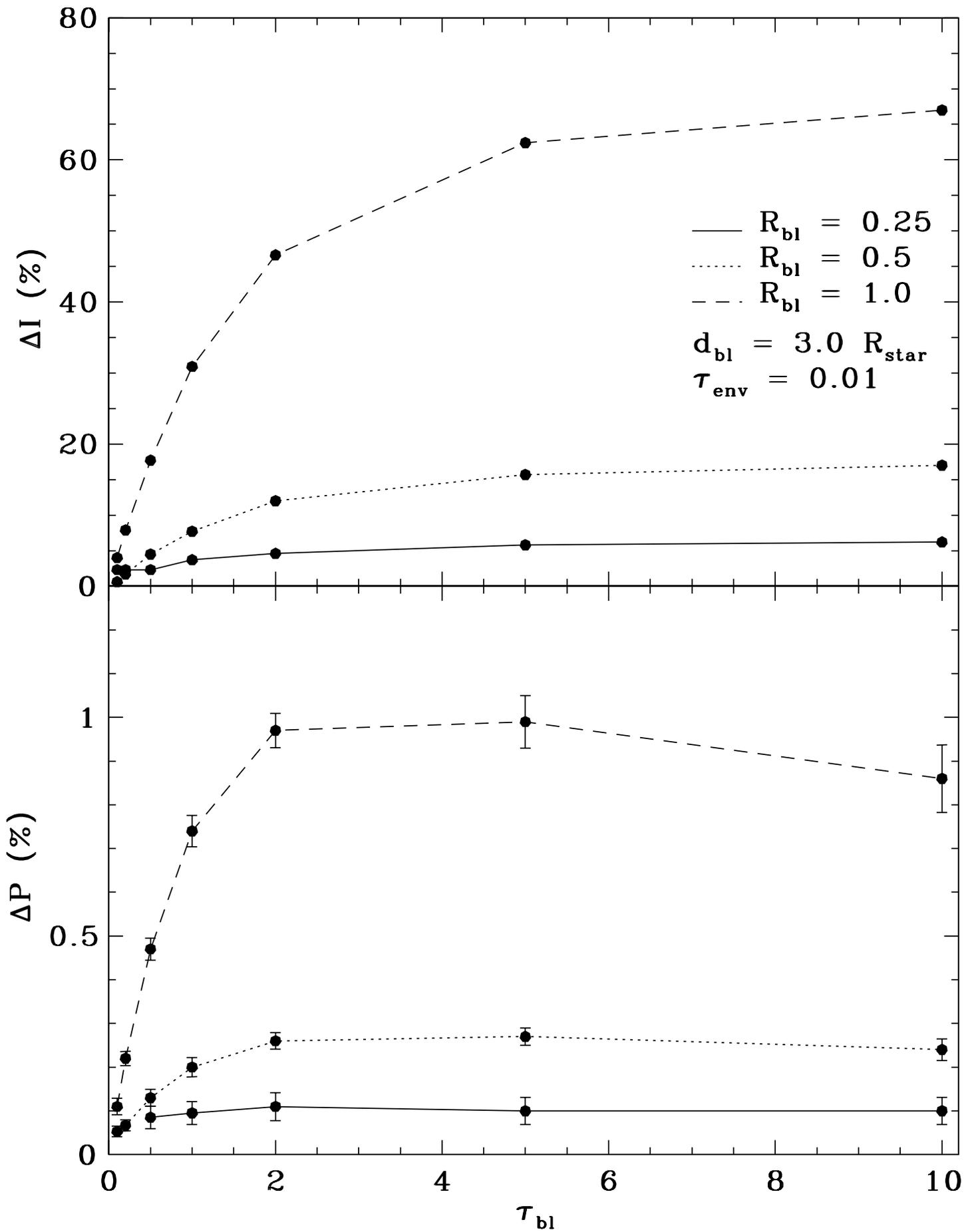
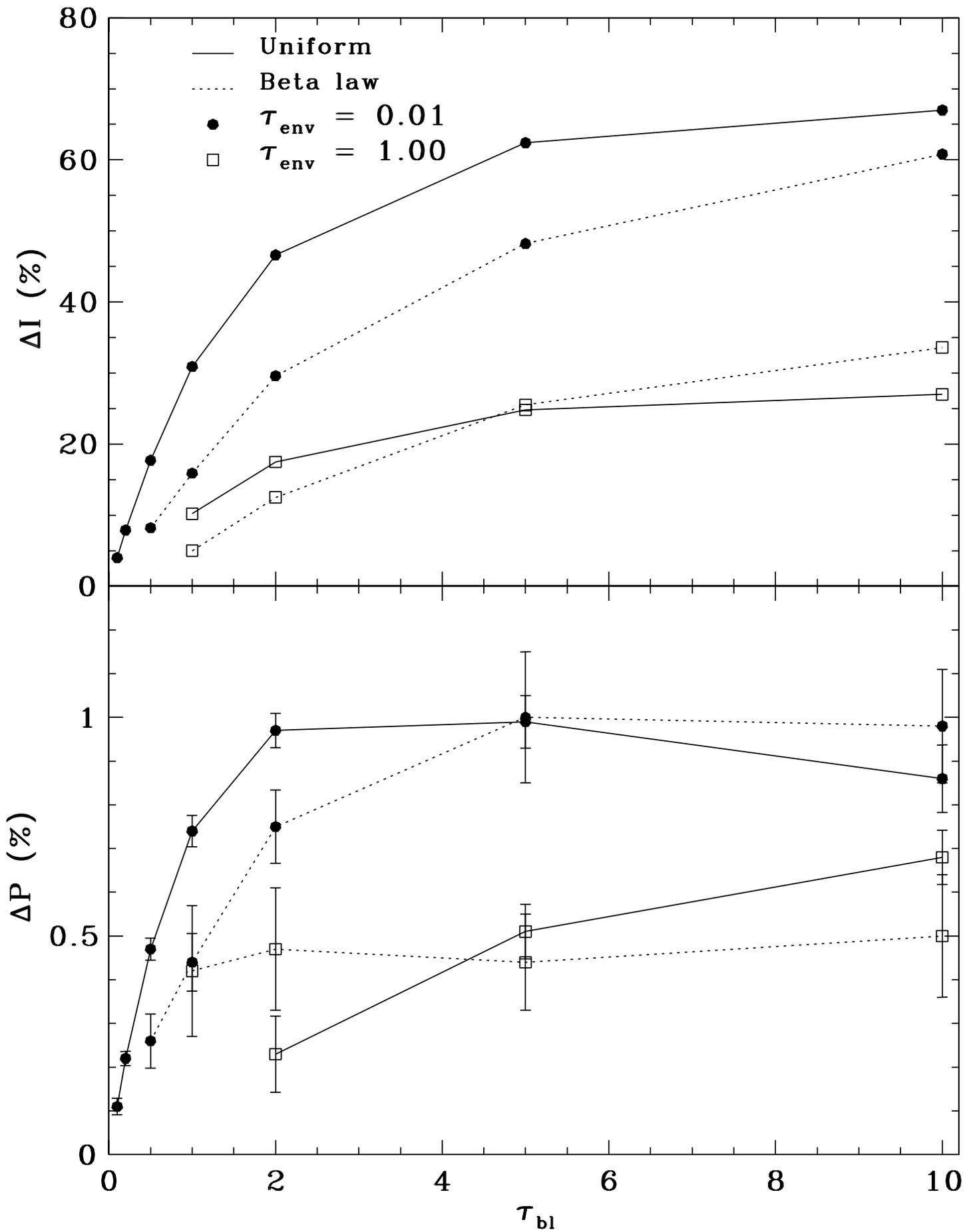
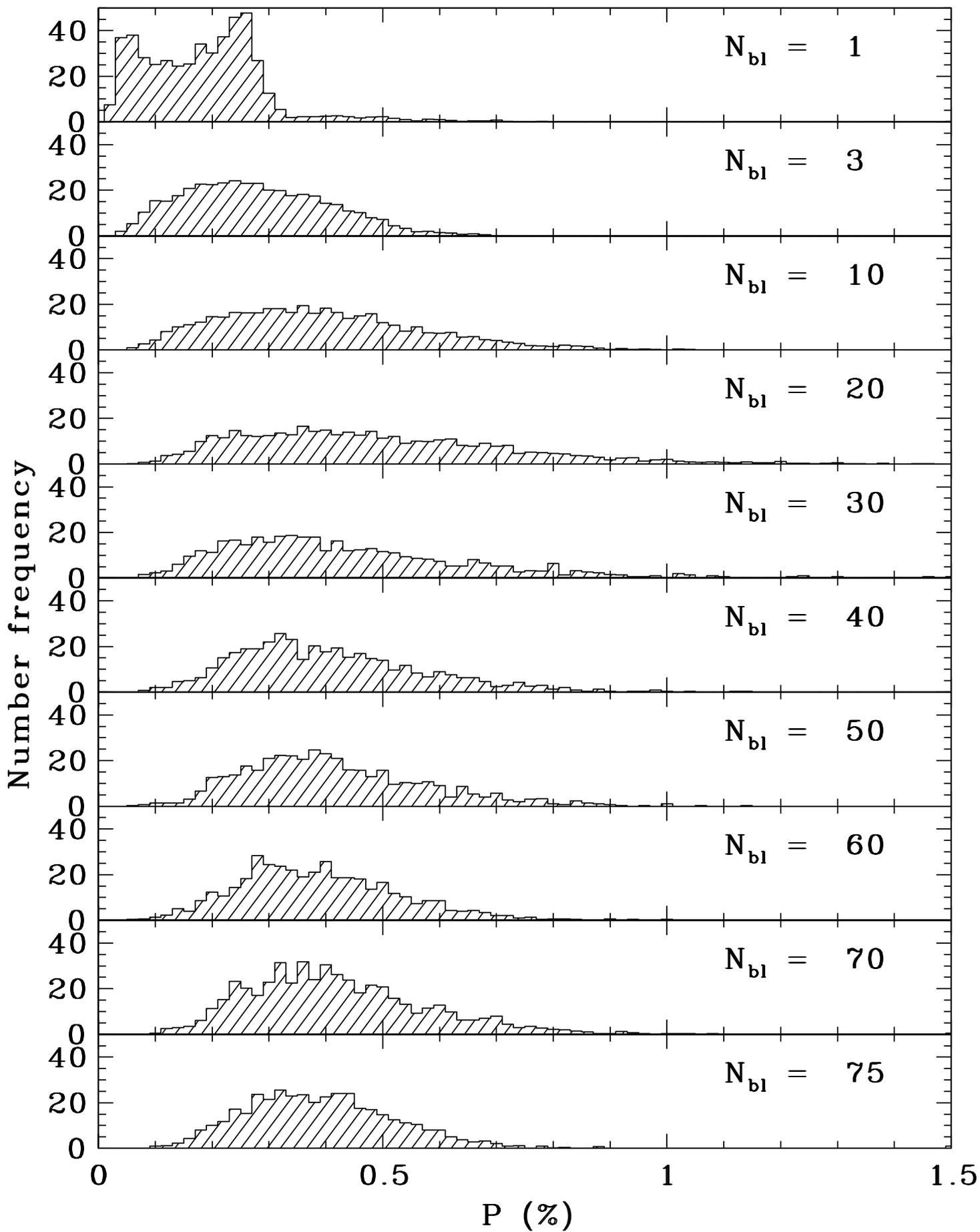
\end{document}